\documentclass[conference]{IEEEtran}
\usepackage[numbers]{natbib}
\usepackage{amsmath}
\usepackage{graphicx}
\usepackage{hyperref}
\usepackage{booktabs}
\usepackage{enumitem}
\usepackage{float}
\usepackage{xcolor}
\usepackage[most]{tcolorbox}
\usepackage{soul}
\usepackage{tikz}

\title{Pick Two: An Adversarial Animal Survival Game}
\author{Jack Vanlyssel, Ramsha Anwar}
\date{}

\begin{document}

\newcommand{\Ramsha}[1]{\textcolor{orange}{Ramsha: #1}}
\newcommand{\jack}[1]{\textcolor{red}{Jack: #1}}

\newtcolorbox{takeawaybox}{
  colback=gray!12,
  colframe=gray!60,
  boxrule=0.5pt,
  arc=2mm,
  left=6pt,
  right=6pt,
  top=6pt,
  bottom=6pt,
  before skip=8pt,
  after skip=10pt,
  fontupper=\small
}

\newenvironment{takeawaylist}{
\begin{itemize}[leftmargin=1.2em,noitemsep,topsep=2pt]
}{
\end{itemize}
}

\maketitle

\begin{abstract}
The “Pick Two” animal selection puzzle is a popular thought experiment in which two animal species must defend a human against the remaining animal attackers \ref{fig:picktwo}. While typically discussed informally, the scenario presents a heterogeneous coalition-selection problem involving complex interactions among agents with different capabilities and behaviors. In this work, we formalize Pick Two as an adversarial multi-agent optimization problem and develop a biologically inspired agent-based simulation framework to evaluate defender coalition effectiveness. Coalition performance is evaluated through 18,000 Monte Carlo simulations conducted in a Unity-based environment. Results show that coalition effectiveness is not additive and is instead dominated by interaction effects and scaling behavior. Overall, this study demonstrates how agent-based simulation can be used to analyze coalition effectiveness in adversarial environments and highlights the importance of emergent group dynamics in determining collective success.

\end{abstract}

\begin{figure}[H]
    \centering
    \includegraphics[width=0.8\linewidth]{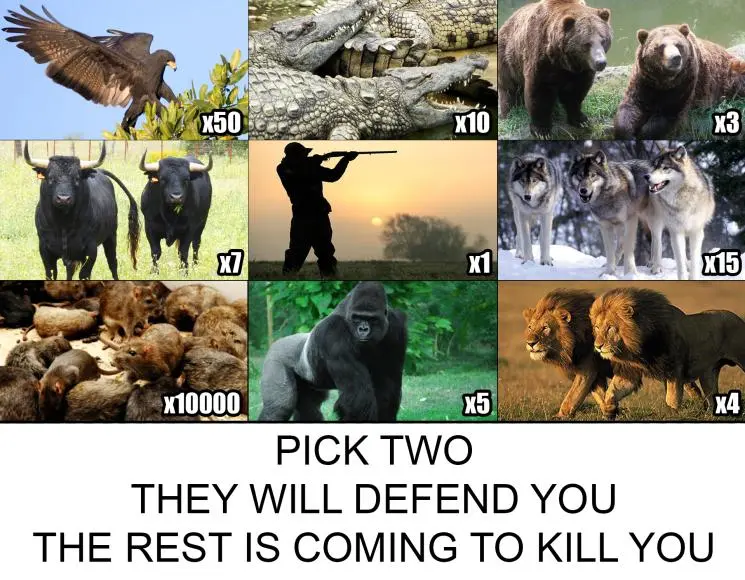}
    \caption{The Pick Two selection puzzle consists of nine animal species. The player must choose two species to defend a VIP against the remaining seven attacking species. The numbers indicate the quantity of animals provided for each species.}
    \label{fig:picktwo}
\end{figure}

\section{Introduction}

The ''Pick Two'' animal selection puzzle is a viral adversarial scenario in which a reader is presented with nine animal species of varying quantity and instructed to select two agents to defend against the remaining seven attackers. Despite its origins as an internet meme, the puzzle raises an interesting strategic question: how does the choice of two defenders influence survival when agents with different capabilities interact in combat?

At face value, Pick Two appears to be a simple toy scenario. However, beneath this informal framing lies a multi-agent problem in which outcomes emerge from complex interactions between heterogeneous agents. Each animal exhibits distinct offensive, defensive, and behavioral characteristics that influence combat differently depending on the opposing composition, spatial positioning, and stochastic progression of the encounter. Consequently, the strongest defensive coalition does not necessarily consist of the two strongest individual animals, but rather the pair whose capabilities interact most effectively under dynamic combat conditions.

This paper formalizes Pick Two as a combinatorial team-selection problem. Each animal is modeled as an agent parameterized using biologically informed attributes derived from biomechanics and ecology. All experiments are conducted within a Unity-based multi-agent simulation environment that incorporates spatial positioning, stochastic combat variability, and heterogeneous behavioral archetypes (e.g., swarm, pack hunter, aerial). Baseline capabilities are first evaluated through one-vs-one encounters to establish relative individual performance under the same simulation conditions. Coalition effectiveness is then evaluated through Monte Carlo trials of the full Pick Two scenario. Performance is measured primarily by win rate, defined as the probability that the selected defender pair survives while protecting a designated human “VIP.” This objective reflects the original puzzle framing in which two animals must defend a person from the remaining attackers. 

We then analyze three core questions:

\begin{enumerate}
    \item How is survival performance distributed across all possible defender pairs?
    \item Do certain coalitions exhibit disproportionately high or low performance relative to the capabilities of their individual components?
    \item How sensitive is performance to parameter adjustment?
\end{enumerate}

To address these questions, we evaluate coalition performance across randomized trials and compare outcomes against baseline individual and group combat behavior observed in one-vs-one encounters. This framework enables the identification of emergent strategies, dominance structures, and interaction effects that are not immediately apparent from individual attributes alone.

We hypothesize that coalition performance is not purely additive and that certain agent pairs will exhibit strong synergistic or antagonistic effects relative to their individual performance. We further hypothesize that performance will be unevenly distributed, with a small subset of agents disproportionately represented in high-performing coalitions.

Although Pick Two originated as an informal thought experiment, it naturally intersects several established research areas, including animal contest theory, agent-based ecological modeling, coalition formation, and adversarial multi-agent systems. As a result, it provides an interesting and accessible testbed for integrating concepts from these domains within a single experimental framework.

\section{Related Work}
This work draws upon concepts from animal contest theory, agent-based ecological modeling, coalition formation, and adversarial multi-agent systems. The following sections review the literature most relevant to modeling animal combat, coalition effectiveness, and strategic behavior in multi-agent environments.

\subsection{Animal Contest Theory}
Animal contest theory provides a framework for understanding how combat outcomes emerge from differences in individual capabilities and strategic behavior. Rooted in evolutionary game theory \cite{smith1973logic}, animal contest theory explains conflict outcomes through Resource Holding Potential (RHP) and behavioral processes such as assessment and escalation \cite{smith1976logic,hardy2013animalcontests}. Recent studies show that both individual and coalition success depend on behavioral and biological factors beyond simple size or numbers \cite{briffa2017role,palaoro2022weapons,anderson2008bite,green2021assessment,green2022fighting,rayner2026collective}. Together, these findings motivate representing animals as heterogeneous agents with distinct physical and behavioral capabilities.

\subsection{Agent-Based Modeling of Animal Systems}
Agent-based models (ABMs) are widely used to study ecological systems composed of interacting individuals. Individual-based ecological models represent organisms as autonomous agents whose local interactions generate population-level behavior \cite{deangelis2014individual}, and have been widely applied to heterogeneous wildlife systems \cite{mclane2011role}. Prior work demonstrates that animal agents can be modeled using internal state, environmental information, and behavioral policies \cite{tang2010agent,deangelis2019decision}. ABMs have also been used to study predator--prey interactions, where complex dynamics and cooperative behaviors emerge from local agent interactions \cite{gras2009individual,wang2020reinforcement}. These results provide a foundation for modeling animal combat as an emergent multi-agent process rather than a predetermined outcome.

\subsection{Coalition Formation and Team Composition}
Coalition formation is a central topic in multi-agent systems research, where teams of agents cooperate to achieve objectives beyond the capability of any individual member \cite{shehory1993coalition}. Research on self-organizing systems shows that complex collective behavior can emerge from local interactions, while coalition formation is often framed as a combinatorial optimization problem \cite{di2005self,kota2012decentralized,sandholm1999coalition}. Studies of heterogeneous teams show that effectiveness depends on interaction effects and synergy rather than individual strength alone \cite{liemhetcharat2012modeling}. These findings suggest that coalition effectiveness cannot be inferred directly from the capabilities of individual members, motivating explicit evaluation of defender pairings.

\subsection{Adversarial Multi-Agent Optimization}
The Pick Two scenario can also be viewed as an adversarial multi-agent optimization problem. Multi-agent systems research provides theoretical foundations for strategic interaction, coordination, and competition among autonomous agents \cite{shoham2002introduction}. Empirical game-theory analysis uses simulation to evaluate strategic behavior in complex environments, while adversarial team games study how coordinated groups optimize against opponents \cite{wellman2025empirical,von1997team}. These studies provide a theoretical basis for viewing Pick Two as an adversarial coalition-selection problem in which defender effectiveness is evaluated against an opposing force.

\subsection{Research Gap}
Prior work has established foundations for modeling animal conflict, simulating ecological systems, forming effective teams, and optimizing strategies in adversarial environments. However, these areas are typically studied in isolation. To our knowledge, no prior work combines these perspectives to investigate how biologically heterogeneous agents form effective defensive coalitions under adversarial pressure. This work addresses that gap by formalizing the Pick Two puzzle as an adversarial coalition-selection problem and evaluating coalition effectiveness within a biologically grounded agent-based simulation framework.
\section{Background}

Before presenting the simulation framework, we review the biological principles underlying the construction of the modeled animal agents. Because the simulation compares species that differ substantially in body size, locomotion, defensive structures, and attack behavior, simplifying assumptions are necessary to map real-world traits to computational parameters. This section summarizes the biomechanical relationships and species traits that inform these mappings.

\subsection{Allometric Scaling in Biomechanics}
\label{Allometry}

\begin{figure}[H]
    \centering
    \includegraphics[width=0.8\linewidth]{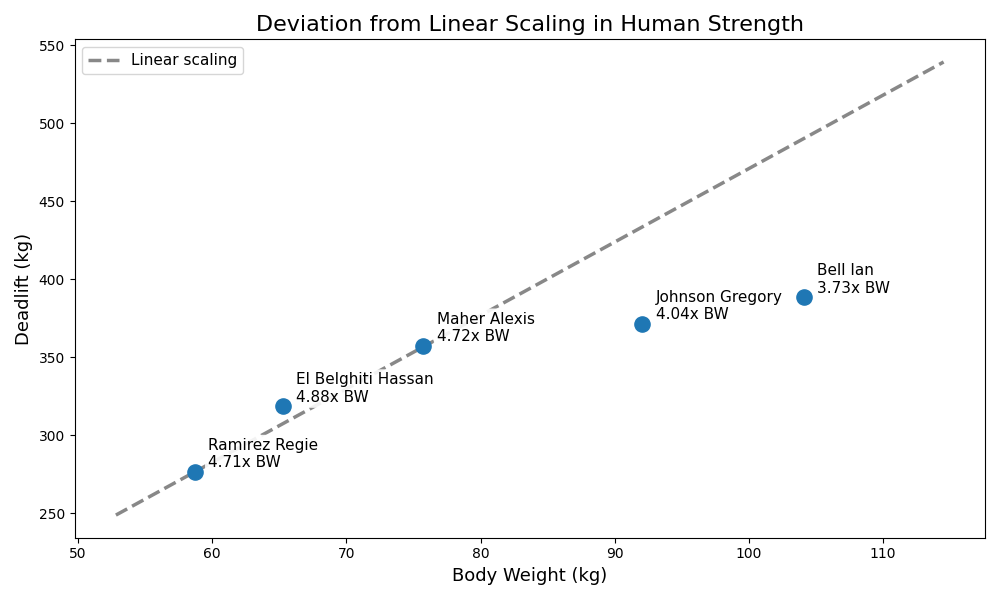}
    \caption{Deadlift strength of world-class powerlifters as a function of body weight with a linear scaling reference. The consistent deviation below the linear trend indicates that strength increases sublinearly with body mass\cite{goodlift_records_2025}.}
    \label{fig:WeightLift}
\end{figure}

Biological performance traits such as strength, durability, and movement ability scale predictably with body mass according to biomechanical allometry, the study of how such traits change systematically with body size across organisms \cite{gittleman2011allometry, shingleton2010allometry} as seen in Figure \ref{fig:WeightLift}.

Under isometric growth, where proportions remain constant, body mass scales with volume ($\propto L^{3}$) while structural strength scales with cross-sectional area ($\propto L^{2}$) \cite{McMahonBodySize, 10.1242/jeb.01520}. This geometric constraint, formalized by Galileo’s square-cube law, implies that strength increases more slowly than mass as visualized in Figure \ref{Allometry}. This produces scaling relationships in which functional performance grows sublinearly with size \cite{galileo1638, haldane_right_size, schmidt-nielsen1984}.

Many biomechanical models express such relationships using power-law scaling \cite{schmidt-nielsen1984, McMahonBodySize} of the form:

\[
Y =\,m^{\alpha},
\]

where \(Y\) represents a biological trait (e.g., strength), \(m\) is body mass, and \(\alpha\) describes how the trait changes with size. Based on the square-cube scaling mentioned above, properties associated with load-bearing area, such as muscle force or skeletal strength, scale with surface area ($L^2$) rather than volume ($L^3$), producing the commonly used prediction \(\alpha \approx 2/3\). We therefore apply this scaling when estimating physical capability across species.

\subsection{Behavioral Differences Across Species}

While biomechanical scaling explains differences in physical capability, interactions between species are also shaped by behavioral strategy. In this work, we model behavioral differences across three dimensions: locomotion, attack mechanics, and coordination. Animals vary in how they move, deliver attacks, and coordinate with others during confrontations, and these differences determine how encounters unfold.

Locomotion influences how animals approach, pursue, and disengage from opponents \cite{wilson2016gsphere}. Species differ in movement strategies, ranging from sustained pursuit to short bursts of acceleration used to close distance and attack \cite{HedenstromScalingSpeed,alexander2003locomotion}. Species differ in attack mechanics due to variation in morphology, weapons, and behavior \cite{palaoro2022weapons}. These differences influence the magnitude and frequency of attacks during combat. Species also differ in how they coordinate during encounters, ranging from solitary behavior to group and swarm-based strategies \cite{hardy2013animalcontests}. These differences can significantly influence combat dynamics.

Together, differences in locomotion, attack mechanics, and coordination determine how animals approach, engage, and disengage during encounters. Capturing these behavioral dimensions is therefore essential for representing realistic interactions between agents. The following section summarizes the biological characteristics of the species included in this study and motivates the parameter choices used in the simulation.

\subsection{Species Overview}
\label{subsec:species-overview}

\begin{table*}[t]
\centering
\begin{tabular}{lllll}
\hline
\textbf{Species} & \textbf{Representative Mass} & \textbf{Top Speed} & \textbf{Primary Attack} & \textbf{Behavior} \\
\hline
\textit{Rattus norvegicus} (Norway rat) & 0.5 kg & 8 mph & Bite & Swarming \\
\textit{Aquila chrysaetos daphanea} (Golden eagle) & 5 kg & 30 mph & Aerial charge & Flying \\
\textit{Canis lupus occidentalis} (Wolf) & 60 kg & 35 mph & Bite & Pack \\
\textit{Homo sapiens} (Human hunter) & 80 kg & 20 mph & Bolt-action rifle & Avoidant \\
\textit{Gorilla beringei graueri} (Gorilla) & 175 kg & 20 mph & Bite, grappling & Bruiser \\
\textit{Panthera leo leo} (Lion) & 190 kg & 30 mph & Bite, grappling & Pack \\
\textit{Crocodylus porosus} (Saltwater crocodile) & 500 kg & 10 mph & Bite & Bruiser \\
\textit{Ursus arctos middendorffi} (Brown bear) & 600 kg & 35 mph & Bite, grappling & Bruiser \\
\textit{Bos taurus} (Toro bravo bull) & 650 kg & 30 mph & Charge & Charger \\
\hline
\end{tabular}
\caption{Representative physical and behavioral characteristics of the animal agents used in the simulation. Behavior corresponds to the archetype used in the agent policy model.}
\label{tab:animal_characteristics}
\end{table*}

Body size varies substantially within many species due to geographic and ecological factors. Consistent with Bergmann’s rule, populations inhabiting colder regions often exhibit larger body sizes than those in warmer environments, producing significant mass variation in widely distributed species such as bears and wolves. Similar variation occurs due to sexual dimorphism (e.g., adult male lion, adult male brown bear, female golden eagle). We model each species using the strongest plausible adult representative rather than population averages that include juveniles or smaller subspecies.

Representative statistics were derived from multiple zoological sources using typical values reported in the literature. For species with substantial geographic variation, values representative of large wild populations were selected rather than global averages that include smaller subspecies. Extreme record individuals were excluded in favor of masses within commonly reported ranges, and when species contained multiple subspecies, a large-bodied representative subspecies was selected.

\subsubsection{\textbf{Norway Rat (Rattus norvegicus)}}
The Norway rat is a small omnivorous rodent known for its agility and adaptability. Adult males typically weigh approximately 0.5~kg \cite{UMichRat, alaska_norway_rat}. Rats exhibit high mobility relative to their size, capable of rapid acceleration and maneuvering through complex environments, with running speeds of approximately 8--10~mph \cite{djawdan1988maximal, garland1983running}. In aggressive encounters, rats rely on rapid repeated biting attacks, with bite forces measured approximately 12~Newtons (N) \cite{Nies2004BiteForce}. They also exhibit a range of social behaviors and can form large aggregations under suitable conditions \cite{schweinfurth2020socialrats,eisenstein1984muricide}.

\subsubsection{\textbf{Golden Eagle (Aquila chrysaetos daphanea)}}
The golden eagle is a large aerial raptor and apex avian predator. Adult females typically weigh approximately 5~kg \cite{pbsEagle,eagles.org,UMichEagle}. Golden Eagles exhibit extremely high aerial mobility, with cruising speeds around 30~mph and dive speeds exceeding 200~mph \cite{UMichEagle}. In predatory encounters, eagles perform high-speed fly-by strikes using their talons \cite{Martinez2014BonelliAttack,Margalida2017EagleBehavior}.

\subsubsection{\textbf{Northwestern Wolf (Canis lupus occidentalis)}}
The gray wolf is a pursuit predator that typically hunts cooperatively in packs. The Northwestern wolf is the largest subspecies of gray wolf, with adult individuals averaging approximately 60~kg \cite{Castello2018Canids, britannica_wolf, UMichWolf}. Wolves exhibit high mobility, with sprint speeds reaching approximately 35--40 mph \cite{mech2003wolves,mech1970wolf}. In aggressive encounters, wolves rely on repeated bites of approximately 1141~N \cite{CarnivoreBiteForce} and exhibit coordinated pack behaviors including flanking and cooperative pursuit \cite{smith2000wolf,macnulty2007wolf}.

\subsubsection{\textbf{Human Hunter (Homo sapiens)}}
The human hunter is modeled as a physically fit adult male with a representative mass of approximately 80~kg \cite{ncd2016height,ncd2017bmi}. Trained humans possess moderate terrestrial mobility, with sprint speeds reaching approximately 20~mph \cite{browning2006walking,Manzer2016KinematicSprint}. In this scenario, the hunter is equipped with a bolt-action rifle chambered in .300~Winchester Magnum, a cartridge commonly used in large-game hunting.

\subsubsection{\textbf{Eastern Lowland Gorilla (Gorilla beringei graueri)}}
The eastern lowland gorilla is the largest living primate and a heavily built terrestrial herbivore with exceptional upper-body strength. Adult males (silverbacks) average approximately 175~kg in mass \cite{UMichgorilla, smithGorillaMass}. Gorillas exhibit moderate terrestrial mobility and can reach short burst speeds of roughly 20~mph, moving primarily through quadrupedal knuckle-walking while occasionally standing or moving bipedally during displays or confrontations \cite{garland1983running, Larson2018PrimateLocomotion}. In aggressive encounters, gorillas rely on grappling, close-range strikes, and powerful bites of approximately 1723~N \cite{GorillaBite,hardy2013animalcontests}.

\subsubsection{\textbf{African Lion (Panthera leo leo)}}
The African lion is a large terrestrial apex predator and one of the most powerful members of the Felidae family. Adult males typically weigh approximately 190~kg \cite{smuts1980lion_growth, UMichLion}. Lions possess strong sprinting capability, reaching speeds of roughly 50--60~km/h (30--40~mph) over short distances when closing on prey, though these bursts can only be sustained briefly \cite{Wilson2018PredatorPreyArmsRace}. Lions frequently operate in coordinated groups and rely on grappling attacks and powerful bites of approximately 2609~N \cite{CarnivoreBiteForce,SmithsonianLionFactSheet,BritannicaLion2026,UMichLion}.

\subsubsection{\textbf{Saltwater Crocodile (Crocodylus porosus)}}
The saltwater crocodile is the largest living reptile and a semi-aquatic ambush predator built for high lethality and extreme durability. To represent a large but typical adult individual while avoiding extreme outliers, we model an adult male with an average mass of approximately 500~kg \cite{webb1989crocodiles,nayak2018crocodile,UMichCroc}. Saltwater crocodiles rely on short bursts of rapid locomotion rather than sustained pursuit, reaching sprint speeds of roughly 10~mph on land over very short distances \cite{BeachSafeCroc}. In aggressive encounters, crocodiles deliver extremely powerful bites (approximately 5792~N) and rely on their large body mass in close-range combat \cite{Erickson2012CrocBite,Ebel2025OsteodermReview}.

\subsubsection{\textbf{Kodiak Brown Bear (Ursus arctos middendorffi)}}
The Kodiak brown bear is one of the largest terrestrial carnivores and, aside from the polar bear, the largest species of bear. Adult males average approximately 600~kg in mass and can reach standing heights of 8--10~ft when upright \cite{UMichBear, britannica_kodiakbear, FWSKodiakBear}. Despite their large body mass, Kodiak bears retain strong short-distance mobility and can reach terrestrial sprint speeds of roughly 35~mph over brief distances \cite{fergus2005bears}. In aggressive encounters, bears rely on powerful forelimbs equipped with large claws and bites of approximately 1627~N \cite{morse2023clawcomparison}.

\subsubsection{\textbf{Spanish Fighting Bull (Bos taurus, toro bravo)}}
The Spanish fighting bull is a large domesticated bovine selectively bred for aggression, endurance under injury, and repeated charging behavior during confrontations. Adult fighting bulls must weigh at least 630~kg by Spanish regulation and typically average around 650~kg \cite{UMichBull, SpanishBullDecree, okstate_fighting_cattle}. Fighting bulls possess strong short-distance mobility and explosive acceleration typical of large bovines, which can reach sprint speeds of approximately 50--56 km/h ($\approx$30--35 mph) \cite{garland1983running}. In aggressive encounters, bulls rely on horn-first charge attacks that combine high momentum with rigid keratin horns to inflict blunt and penetrating trauma \cite{lomillos2021hornstructure}.

\section{Methodology}
To evaluate defensive animal pairings under controlled and repeatable conditions, we developed a physics-based multi-agent combat simulator. Because direct experimental comparisons between species are impractical and unethical, the simulation provides a structured framework for exploring how differences in body mass, mobility, attack mechanics, and group behavior influence combat outcomes. The following sections describe the simulation environment, agent representation, combat mechanics, behavioral policies, and experimental protocol used in the study.

\subsection{Simulation Overview}
\begin{figure}[H]
    \centering
    \includegraphics[width=0.8\linewidth]{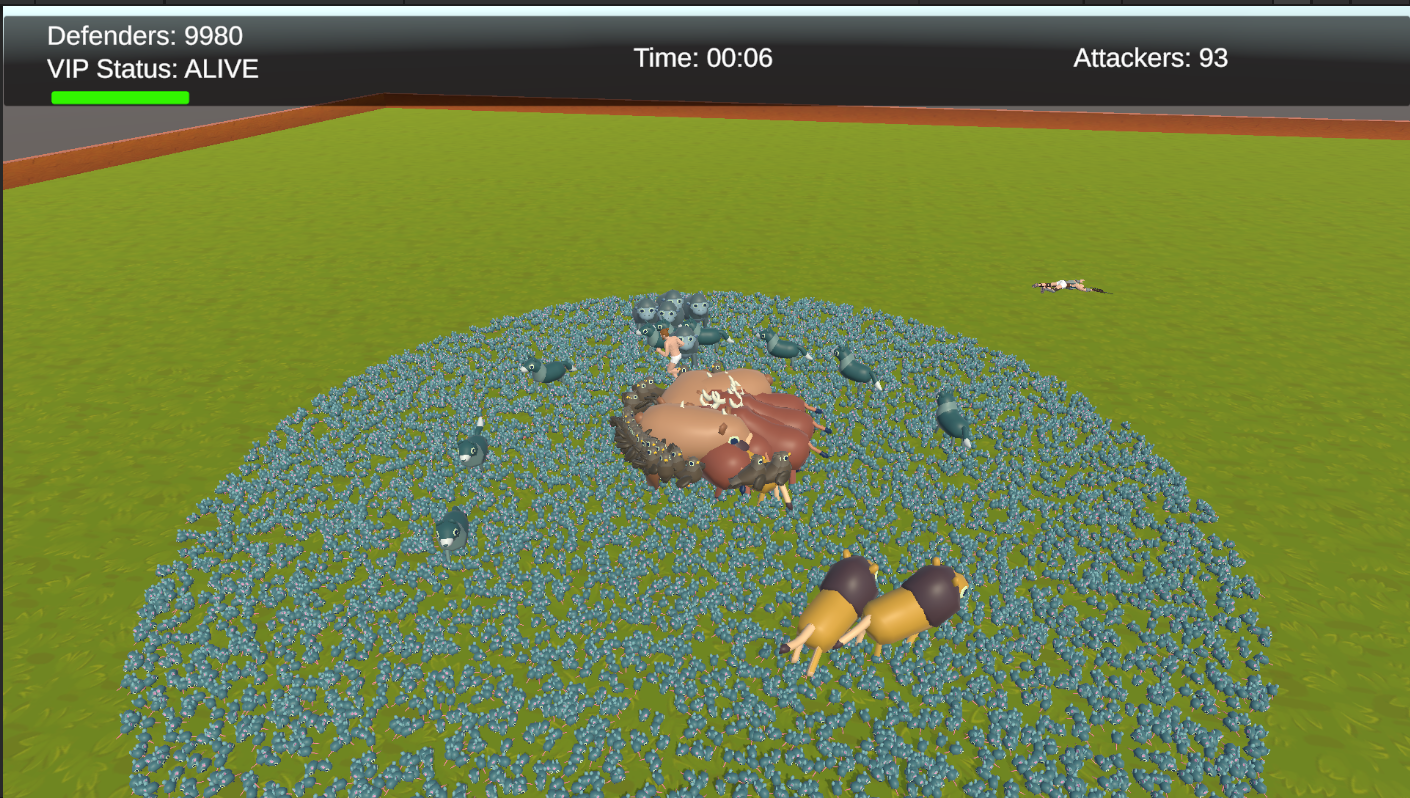}
    \caption{A swarm of rats engaging other animals in the Unity simulation}
    \label{fig:Sim Pic}
\end{figure}

We implement our combat simulator in the Unity game engine (seen in Figure \ref{fig:Sim Pic}). Each trial initializes a set of animal agents and a human VIP agent in a bounded environment. Agents interact through movement and attack actions. Each agent features a state controller that governs transitions between behavioral states (e.g., idle, chase, attack, retreat) based on environmental conditions and internal behavior variables. Each trial terminates when all attackers or defenders are neutralized or when a simulation time limit of 300 seconds is reached. The simulator updates agent behaviors once per frame using Unity's real-time update loop, with time-dependent processes such as movement and attacks scaled by the frame interval ($\Delta t$) to ensure consistent behavior across varying frame rates.

\subsection{Agent Representation}
Each animal species is modeled as an agent with attributes inspired by its biology (Section~\ref{subsec:species-overview}). These attributes include:
(i) defensive attributes (Hit Points (HP) and armor),
(ii) an attack behavior defining damage, range, and attack cadence,
(iii) mobility attributes (speed and turning rate), and
(iv) a behavior policy (archetype).

In addition to these attributes, each agent is assigned a geometric representation used for interaction. Because the rendered animal models are stylized for visual clarity, mesh size is not treated as a physically meaningful quantity in the simulator. 

\subsection{Defensive Model: HP and Armor}
\label{DefensiveModel}

\begin{table}[htbp]
\centering
\begin{tabular}{lccc}
\hline
\textbf{Species} & \textbf{HP} & \textbf{Armor} & \textbf{Defensive Features} \\
\hline
Rat & 1.00 & None (0.0) & Minimal tissue protection \\
Eagle & 4.68 & None (0.0) & Light frame, feathers \\
Wolf & 24.72 & Low (0.1) & Fur, moderate muscle \\
Hunter & 29.98 & None (0.0) & Minimal tissue protection \\
Gorilla & 50.64 & Medium (0.2) & Thick muscle, dense bone \\
Lion & 53.51 & Low (0.1) & Fur, muscle shielding \\
Crocodile & 102.33 & High (0.4) & Osteoderms, thick hide \\
Bear & 115.62 & High (0.4) & Fat, fur, heavy muscle \\
Bull & 122.00 & Medium (0.2) & Thick hide, muscle mass \\
\hline
\end{tabular}
\caption{Defensive attributes for each species. HP is derived from mass using Eq. (HP scaling). Armor represents multiplicative damage reduction with discrete levels $A \in \{0.0, 0.1, 0.2, 0.4\}$.}
\label{tab:defensive_attributes}
\end{table}

\subsubsection{HP Scaling}
HP is interpreted as an abstract durability metric representing an agent's capacity to sustain damage before incapacitation. It does not correspond directly to physiological health or wound severity, but instead serves as a computational threshold for agent removal. To preserve biologically plausible durability while avoiding unrealistic invulnerability in large species, HP scales allometrically with body mass:
\[
HP(w) = HP_{\text{rat}}\left(\frac{w}{w_{\text{rat}}}\right)^{\alpha},
\]
where the smallest animal, $HP_{\text{rat}} = 1$, anchors durability. We set $\alpha = 2/3$ following biomechanical square–cube scaling relationships discussed in Section \ref{Allometry} \cite{haldane_right_size, schmidt-nielsen1984}. This ensures that larger animals are more durable while preserving diminishing returns consistent with biomechanical allometry \cite{schmidt-nielsen1984}.

\subsubsection{Armor}
Some animals possess substantially greater passive protection than others, resulting in reduced damage from incoming attacks. We model this effect using an armor coefficient that captures damage mitigation arising from skin thickness, hide density, fur/fat layers, muscle mass, and dermal armor. Damage reduction is applied multiplicatively:
\[
D_{\text{applied}} = D_{\text{raw}}(1 - A),
\]
where \(A \in [0,1)\) is the armor coefficient. We define four discrete armor tiers:
\[
A \in \{0.0, 0.1, 0.2, 0.4\} \quad \text{(None, Low, Medium, High)}.
\]
These tiers provide progressively increasing levels of protection while remaining bounded such that even the highest tier does not fully trivialize incoming damage. This ensures that armor acts as a secondary modifier rather than dominating the HP-based durability model.

Because cross-species quantitative measurements of damage reduction are not directly available, armor values are treated as modeling assumptions. Tier assignments were made using observable biological characteristics, including:

\begin{itemize}[noitemsep, topsep=2pt]
\item Skin and hide thickness
\item Presence of dense fur, fat, or muscle shielding
\item Structural protection (e.g., skull robustness, osteoderms)
\item Known resilience to physical trauma in studies
\end{itemize}

This approach provides a coarse but biologically grounded differentiation between lightly protected and heavily protected species while maintaining model simplicity. Both HP and armor values for each agent can be seen in Table \ref{tab:defensive_attributes}.

\subsection{Offensive Model: Attacks and Damage}

\begin{table}[htbp]
\centering
\begin{tabular}{lcccc}
\hline
\textbf{Species} & \textbf{Range} & \textbf{Attack Rate (1/s)} & \textbf{Damage} & \textbf{Special Case} \\
\hline
Rat & 1 & 0.5 & 0.359 & Swarm \\
Eagle & 1 & 0.1 & 0.974 & Aerial, Charge \\
Wolf & 2 & 0.5 & 7.495 & -- \\
Hunter & 20 & 1 & 25 & Ranged \\
Gorilla & 3 & 1 & 9.865 & -- \\
Lion & 3 & 1 & 13.008 & -- \\
Crocodile & 3 & 0.5 & 22.137 & -- \\
Bear & 3 & 1 & 9.495 & --\\
Bull & 3 & 0.1 & 25 & Charge \\
\hline
\end{tabular}
\caption{Offensive attributes for each species. Damage values are derived from class-specific anchors (wolf for close-range, bull for charge, and firearm for ranged). Attack rate reflects combat style.}
\label{tab:offensive_attributes}
\end{table}

Each agent implements offensive behavior through an attack model defined by attack range \(r\), base damage \(D_{\text{raw}}\), and attack cooldown. 

An attack represents a discrete damage event in which an agent successfully inflicts damage on a target. The frequency of attacks is controlled by the agent's attack rate, which abstracts differences in fighting style across species. For some agents, such as wolves or bears, an attack event represents a short interval of close-range combat, while for charging or ranged agents such as bulls, eagles, and the hunter, it corresponds to a single high-impact action. This abstraction enables diverse attack mechanisms to be represented within a common framework.

\subsubsection{Attack Types}
To maintain consistency across species while capturing different combat mechanisms, we define three primary attack classes:

\begin{itemize}[noitemsep, topsep=2pt]
\item \textbf{Close-Range Attacks:} Repeated close-range attacks (e.g., wolf, lion, bear, gorilla) delivered during sustained engagement.
\item \textbf{Charge Attacks:} High-momentum, burst-based attacks (e.g., bull, eagle) that deliver large damage in a single event.
\item \textbf{Ranged Attacks:} Distance-based attacks delivered from outside the immediate contact range (e.g., hunter), providing high-impact damage.
\end{itemize}

These classes reflect the dominant modes of force delivery observed across the modeled species and allow damage to be anchored to physically interpretable reference cases.

\subsubsection{Incapacitation-Based Damage Calibration}
Because cross-species measurements of damage are not comparable, each attack class is anchored to a representative real-world reference.

\textbf{Close-Range Anchor (Wolf):}  
Documented wolf attacks on humans suggest that a single successful attack is often serious but not usually fatal, whereas repeated successful attacks can become incapacitating \cite{ColoradoState_wolf_safety, Linnell2021WolfAttacks, Ambarli2019WolfConflict}. To reflect this severity range while preserving survivability under brief contact, baseline wolf damage was calibrated so that an average human would be incapacitated after approximately four full-contact attacks. With human HP set to 29.98, this yields:

\[
D_{\text{wolf}} = 7.495
\]

Damage for other close-range attacks is then scaled relative to the wolf using a sublinear function of bite force:

\[
D_i = D_{\text{wolf}} \left(\frac{BF_i}{BF_{\text{wolf}}}\right)^{2/3},
\]

where \(BF_i\) is the bite force of species \(i\). The exponent \(2/3\) introduces diminishing returns so that extreme differences in raw bite force do not translate linearly into damage, reflecting biomechanical constraints on how force translates to effective injury.

\textbf{Charge Anchor (Bull):}  
Documented bull attacks on humans frequently result in severe blunt trauma, goring injuries, or trampling and are often capable of incapacitating a victim in a single high-momentum interaction \cite{Byard2024CattleForensic, Dogan2008BullAttacks, CDC2009CattleFatalities}. To reflect this lethality while preserving consistency with the human durability model, baseline charge damage was calibrated such that a full-force impact corresponds to near human incapacitation. With human HP set to 29.98, this yields:

\[
D_{\text{bull}} = 25
\]

Damage for other charge-based attacks is then scaled relative to the bull using a sublinear function of impact momentum:

\[
D_{\text{charge}} = D_{\text{bull}} \left(\frac{m_i v_i}{m_{\text{bull}} v_{\text{bull}}}\right)^{2/3},
\]

where \(m_i\) and \(v_i\) are the mass and sprint speed of species \(i\). The exponent \(2/3\) introduces diminishing returns so that extreme differences in momentum do not translate linearly into damage, reflecting biomechanical and collision constraints on how impact energy is transferred during large-body interactions.

\textbf{Ranged Attack (Hunter):}  
The human hunter is equipped with a bolt-action rifle in the .300 Winchester Magnum class, a high-powered weapon used for large-game hunting. It provides accurate long-range engagement and high-impact energy. The weapon is modeled as a ranged, high-precision attack with a relatively long reload time, reflecting bolt-action cycling. Due to its high energy density and penetrative capability, rifle fire is assumed to bypass biological armor and deliver consistent damage on a successful hit. To balance lethality while preserving the possibility of survival, rifle damage is set to:

\[
D_{\text{gun}} = 25
\]

This value reflects a near-lethal strike against a human target, capturing the severe but not always instantaneous incapacitation associated with firearm trauma.

\textbf{Attack rate} defines how frequently an agent generates damage opportunities and is assigned according to fighting style. Repeated bite attackers (e.g., wolves, crocodiles) operate at 0.5 attacks/s (2 s cooldown), reflecting intermittent single-point attacks. Close-range fighters capable of applying force through multiple appendages simultaneously (e.g., bears, lions, gorillas) operate at 1.0 attacks/s. Burst-based charge attackers (e.g., bulls, eagles) operate at 0.1 attacks/s (10 s cooldown), reflecting infrequent but high-magnitude engagements. This formulation captures trade-offs between attack frequency, attack delivery style, and per-attack magnitude, while maintaining a consistent, interpretable combat tempo.

\textbf{Attack range} defines the distance at which an agent can successfully apply damage to a target. Species are grouped by mass as small ($<10$ kg), medium ($10$--$150$ kg), and large ($>150$ kg), with ranges of 1, 2, and 3 Unity units (1 unit = 1 meter), respectively; the hunter is a long-range exception with a range of 20. Attack eligibility is determined using center-to-center distance checks and species-specific attack ranges, with damage applied when a valid target is within range and the attack is off cooldown.

\subsection{Mobility Model}

\begin{table}[htbp]
\centering
\small
\setlength{\tabcolsep}{4pt}
\begin{tabular}{lccc}
\hline
\textbf{Species} & \textbf{Sprint (m/s)} & \textbf{Turn (°/s)} \\
\hline
Rat & 3.6 & 700 \\
Eagle & 13.4 & 700 \\
Wolf & 15.6 & 500 \\
Hunter & 8.9 & 500 \\
Gorilla & 8.9 & 300 \\
Lion & 13.4 & 300 \\
Crocodile & 4.5 & 300 \\
Bear & 15.6 & 300 \\
Bull & 13.4 & 300 \\
\hline
\end{tabular}
\caption{Mobility parameters for each species. Movement is defined by sprint speed and turning rate, which captures maneuverability differences across body sizes.}
\label{tab:mobility}
\end{table}

Agent mobility is defined by sprint speed and maneuverability. Sprint speed was anchored to approximate maximum speeds reported for each species (Section~\ref{subsec:species-overview}) and converted to Unity units using a scale of 1 unit = 1 meter. All agents move at their assigned sprint speed during combat.

Turning rate serves as a proxy for maneuverability and is assigned based on body size. Smaller animals exhibit higher turning rates due to lower inertia, while larger animals are more constrained. Species were grouped into three mass-based classes ($<10$ kg, $10$--$150$ kg, $>150$ kg), corresponding to turning rates of 700$^\circ$/s, 500$^\circ$/s, and 300$^\circ$/s, respectively \cite{howland1974predatoravoidance, moore2015outrun}.

\subsection{Behavior Policies (Archetypes)}
Because physical attributes alone do not fully determine combat behavior, agents must also exhibit appropriate decision-making (e.g., a hunter maintaining distance when using a ranged weapon, or wolves operating as a group). To capture these differences without introducing excessive complexity, we represent behavior using a small set of simple archetypes that reflect common combat strategies observed in nature. This abstraction ensures consistent and comparable decision-making across species, capturing key behaviors such as pursuit and group coordination.

We implement species behavior as the following archetypes:

\begin{itemize}
  \item \textbf{Bruiser:} acquires nearby targets, closes to attack range, and sustains close-quarters combat; may disengage/reposition based on target validity (bear, gorilla, crocodile, rat).
  \item \textbf{Avoidant:} the agent retreats when targets are too close and engages with ranged attacks from a safer distance (human).
  \item \textbf{Charger:} opportunistically enters a charge state running through enemies when cooldown conditions permit and the target is valid (bull).
  \item \textbf{Pack:} follows a shared target, orbiting and then attacking based on cooldown (wolf, lion).
  \item \textbf{Aerial:} approaches target and charges based on cooldown. Only targetable when in the charge, not while moving normally due to flight (eagle).
\end{itemize}

Policies are implemented as finite-state machines that govern target selection, movement, engagement, and disengagement behavior according to the rules of each archetype.

\subsection{Agent Aggregation}
Some species are represented by extremely large populations, making explicit simulation of every individual agent computationally prohibitive. To maintain tractable large-scale simulations while approximating collective behavior, we use an aggregated swarm model in which many individuals are abstracted into a single composite agent. Although currently used only for rats, this framework generalizes to other swarm-like organisms or highly numerous agents.

Swarm HP is determined by multiplying the number of represented units by individual HP, producing a shared health pool. Incoming damage proportionally reduces the living population, while offensive capability scales with the number of surviving members. Swarm damage is implemented as a periodic damage-over-time (DOT) system rather than discrete attack events. At fixed intervals, the swarm evaluates nearby targets, allocates finite members across those targets, and increases the number of engaged members over time. Engaged members are tracked separately per target to prevent individual units from simultaneously attacking multiple enemies. Engagement capacity scales sublinearly with target mass, such that larger animals can be attacked by more swarm members. 
Engagement capacity scales sublinearly with target mass to reflect the fact that larger animals provide more attack surface but not in direct proportion to body mass.
Crucially, the target can attack the swarm to remove engaged agents. Damage per tick equals the number of engaged members multiplied by the per-member damage value.

The swarm’s visual representation is also abstracted for performance reasons, rendering a central model surrounded by clones with no animations. Together, these mechanisms provide a computationally efficient approximation of collective attack behavior.

\subsection{Environment and Trial Setup}
Trials are conducted in a flat, bounded arena measuring 150 × 150 Unity units. The VIP and two defending agents are initialized at the center of the arena. Attacking agents are spawned at a fixed radius of 60 units from the VIP, evenly spaced in randomized positions along a surrounding circle to ensure equal initial distance while allowing variation in approach direction. All stochastic elements are generated from a deterministic pseudorandom seed assigned to each trial. The seed controls attacker spawn orientation, attacker placement, overlap-resolution behavior, and other stochastic simulation events.

The VIP's objective is survival. Attackers attempt to engage and incapacitate defenders and the VIP, while defenders attempt to intercept and neutralize incoming threats. Unlike combat agents, the VIP does not attack and instead follows a simple evasive movement policy. At each update, the VIP moves away from nearby attackers while remaining near surviving defenders, allowing limited self-preservation behavior without introducing complex decision-making.

Attackers prioritize defenders while at least one non-aerial defender remains alive. During this phase, the VIP is not directly targeted. Once all defenders are neutralized, attackers re-target and engage the VIP directly. Each trial terminates when one of the following conditions is met: (i) all attackers are neutralized, (ii) the VIP is incapacitated, or (iii) a simulation time limit of 300 seconds is reached.

This design isolates defender effectiveness by ensuring that attackers must first contend with defenders before engaging the VIP, while allowing the VIP to exhibit limited evasive behavior.

\subsection{Calibration}
Prior to multi-agent experiments, we conducted one-vs-one encounters between species to validate baseline combat dynamics. Each trial consisted of two opposing species in a flat, bounded arena. We evaluated both individual encounters involving a single representative of each species and group encounters using the species population counts defined in the original Pick Two scenario.

These trials were used to confirm that the model produced qualitatively plausible interaction patterns (e.g., larger predators outperforming smaller animals under direct engagement). Calibration was limited to ensuring internal consistency and stability of outcomes; no parameters were tuned to "balance" multi-agent performance. The resulting parameter set was then fixed and used without modification in all subsequent experiments.

\subsection{Experimental Protocol}
Each defender pairing was evaluated across 500 Monte Carlo trials. With nine species, this yields 36 unique defender pairings and a total of 18,000 simulated battles (36 pairings × 500 trials). The number of trials was chosen to ensure stable win-rate estimates while remaining computationally tractable.

All simulations were executed on a desktop system (Windows 11, Intel Core i7-13700K CPU, 32 GB RAM) using up to 24 parallel instances of the compiled program at normal simulation speed. This avoids artifacts introduced by time scaling in timing and physics-dependent interactions. Each trial is assigned a deterministic pseudorandom seed, and all stochastic decisions within the simulation are derived from that seed. Randomness arises from spawn orientation, attacker placement, target selection, overlap resolution behavior, and minor behavioral timing variation.

Independent variables include defender pair selection and seed, while dependent variables include win rate (VIP survival), time to termination, number of surviving defenders, and per-agent damage contribution.

\subsection{Data Analysis}
Simulation data from each bout was exported as structured output logs and analyzed using custom Python scripts. Recorded variables included trial outcome, time to termination, surviving agents, per-agent damage contribution, and other combat statistics used for post-processing and aggregation. The win rate for each defender pair was estimated as the fraction of trials in which the VIP survived. Results from one-vs-one, group, and Pick Two experiments were then aggregated and compared to identify patterns in species and coalition performance.
\section{Results}
To frame our results, we first examine biological factors such as total biomass and the number of units, and predict combat outcomes. Next, we present one-vs-one statistics to establish baseline combat relationships between individual species. We then evaluate the full Pick Two scenarios and finish with a sensitivity analysis to see how altering the proportions of each species changes outcomes.

\subsection{Predictions}
We first consider simple predictors of dominance based on biomass and unit count. Under a biomass-only model, species with the greatest total mass would be expected to dominate. Rats and crocodiles stand out, each contributing 26\% of the total biomass, suggesting they should perform similarly well.

However, biomass alone is not enough to predict outcomes. Species with many smaller units, such as rats, can cause more damage more frequently because more individuals can attack at once. In addition, allometric scaling means larger animals gain less effective strength per unit mass. Together, this suggests that high-count species may outperform equally massive but less numerous opponents. You can see species rankings based on these stats in Table \ref{tab:biomass_ranking}:

\begin{table}[H]
\centering
\caption{Species ranked by total biomass (count $\times$ mass). Percentages represent the share of total biomass (19,215 kg).}
\label{tab:biomass_ranking}
\begin{tabular}{lcccc}
\toprule
\textbf{Rank} & \textbf{Species} & \textbf{Count} & \textbf{Biomass (kg)} & \textbf{Share (\%)} \\
\midrule
1 & Rat & 10{,}000 & 5{,}000 & 26.0 \\
2 & Crocodile & 10 & 5{,}000 & 26.0 \\
3 & Bull & 7 & 4{,}550 & 23.7 \\
4 & Bear & 3 & 1{,}800 & 9.4 \\
5 & Wolf & 15 & 900 & 4.7 \\
6 & Gorilla & 5 & 875 & 4.6 \\
7 & Lion & 4 & 760 & 4.0 \\
8 & Eagle & 50 & 250 & 1.3 \\
9 & Human & 1 & 80 & 0.4 \\
\bottomrule
\end{tabular}
\end{table}

\subsection{One-vs-One Results}
\textbf{Single Unit Trials: } To establish baseline combat relationships between species, we first examine one-vs-one interactions. These pairwise encounters isolate individual capability by removing group effects such as coordination, swarm dynamics, and numerical advantage. As a result, outcomes in this setting primarily reflect differences in per-unit attributes, including durability, damage, and attack cadence.

\begin{figure}[H]
    \centering
    \includegraphics[width=0.8\linewidth]{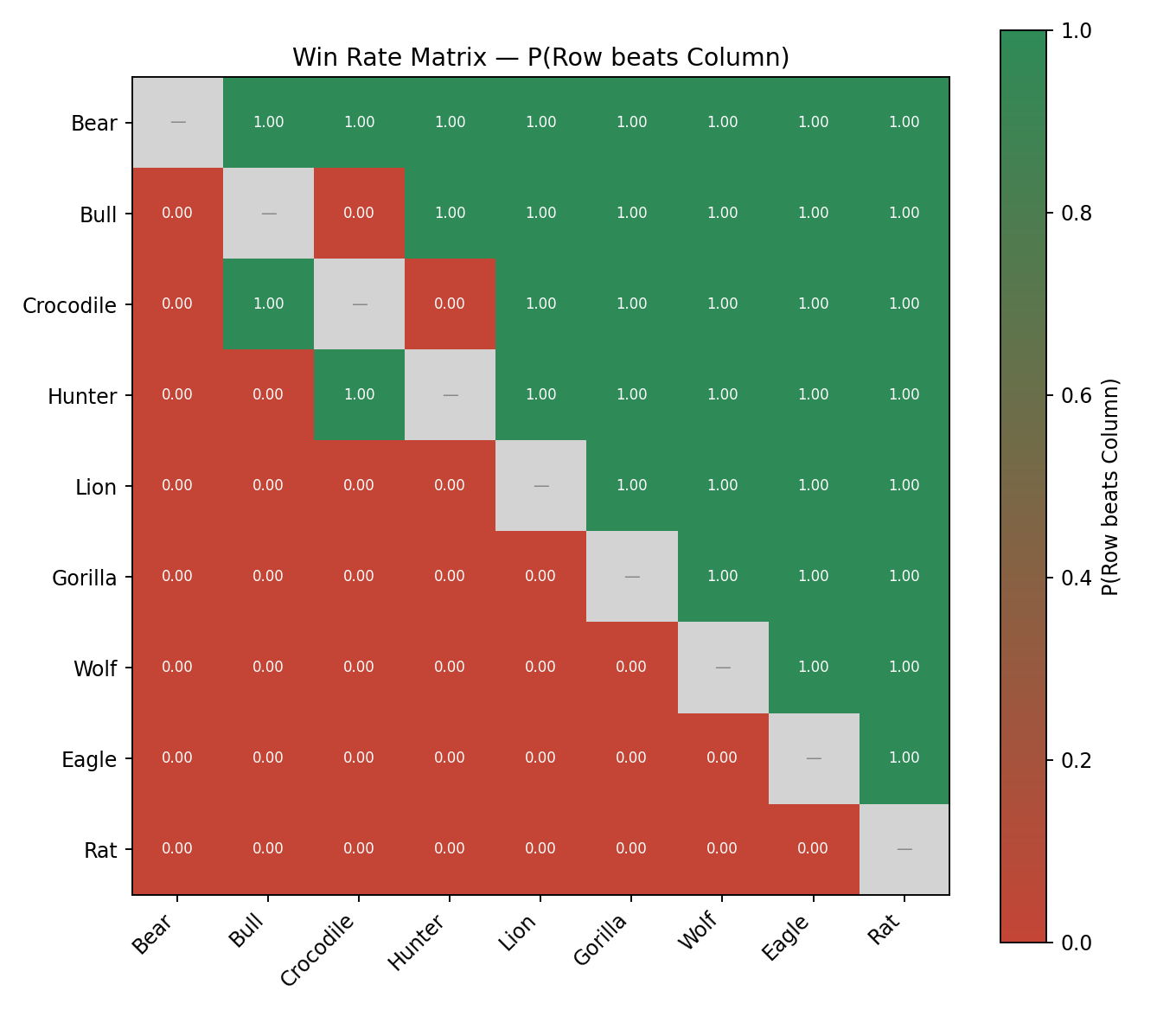}
    \caption{Pairwise win-rate matrix for one-vs-one encounters. Each cell represents the probability that the row species defeats the column species, estimated over 500 Monte Carlo trials per matchup.}
    \label{fig:solo_win_rate}
\end{figure}

Larger animals generally outperform smaller ones due to higher HP and per-attack damage (shown in Figure \ref{fig:solo_win_rate}), but the results reveal important deviations from simple mass-based scaling. Bears outperform bulls despite lower mass due to higher attack rates and armor. Crocodiles are also lower mass than bulls, but have more damage per strike. The hunter’s high single-target damage and range make him highly effective in one-vs-one, allowing him to eliminate opponents before being overwhelmed. Smaller species such as eagles and rats perform worst overall, resulting in a largely consistent but not strictly mass-ordered hierarchy.

These deviations reflect the influence of attack type, mobility, and special behaviors beyond size alone. The resulting one-vs-one outcomes establish a baseline ranking of individual strength. The consistency of these results provides validation for the underlying model, allowing us to proceed to group-based trials.

\begin{figure}[H]
    \centering
    \includegraphics[width=0.8\linewidth]{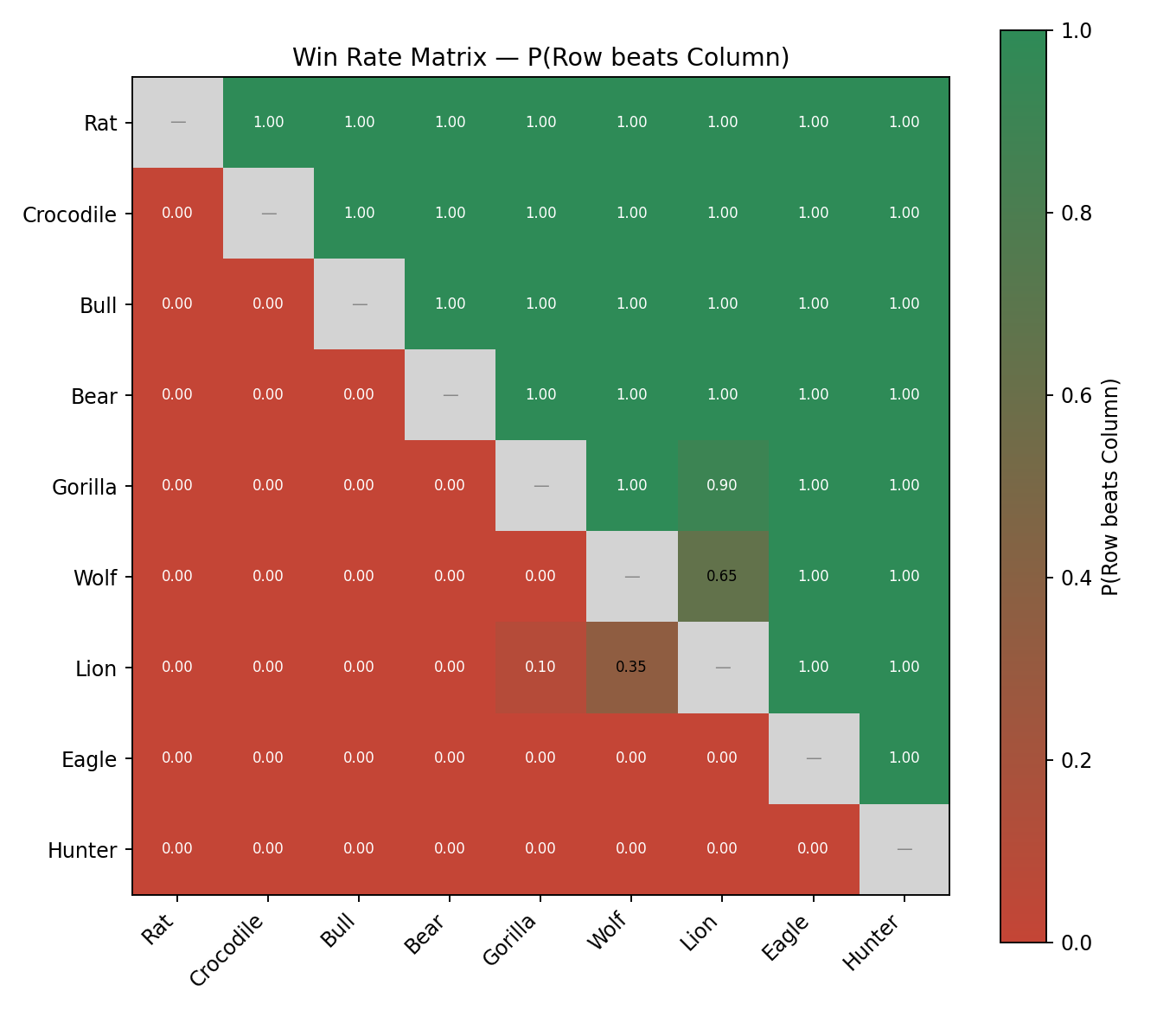}
    \caption{Pairwise win-rate matrix for group one-vs-one encounters. Each cell represents the probability that the row species defeats the column species, estimated over 500 Monte Carlo trials per matchup.}
    \label{fig:group_win_rate}
\end{figure}

\textbf{Group Unit Trials:} To evaluate how individual performance scales in multi-agent settings, we next examine group-based one-vs-one encounters. These trials introduce interactions such as coordination and collective behavior, allowing us to assess how species perform when multiple units are present. Outcomes in this setting reflect not only per-unit attributes but also how effectively those attributes scale under group dynamics. These results form our baseline stats for each species that are compared with Pick Two performance.

The transition from single-unit to group-based one-vs-one encounters produces a markedly different set of outcomes. As shown in Figure~\ref{fig:group_win_rate}, the win-rate matrix remains highly polarized, with most matchups converging to near-deterministic outcomes. This indicates that in multi-unit settings, small per-unit advantages compound, eliminating much of the stochastic variability observed in isolated encounters and leading to decisive dominance rather than balanced contests.

\begin{figure}
    \centering
    \includegraphics[width=0.8\linewidth]{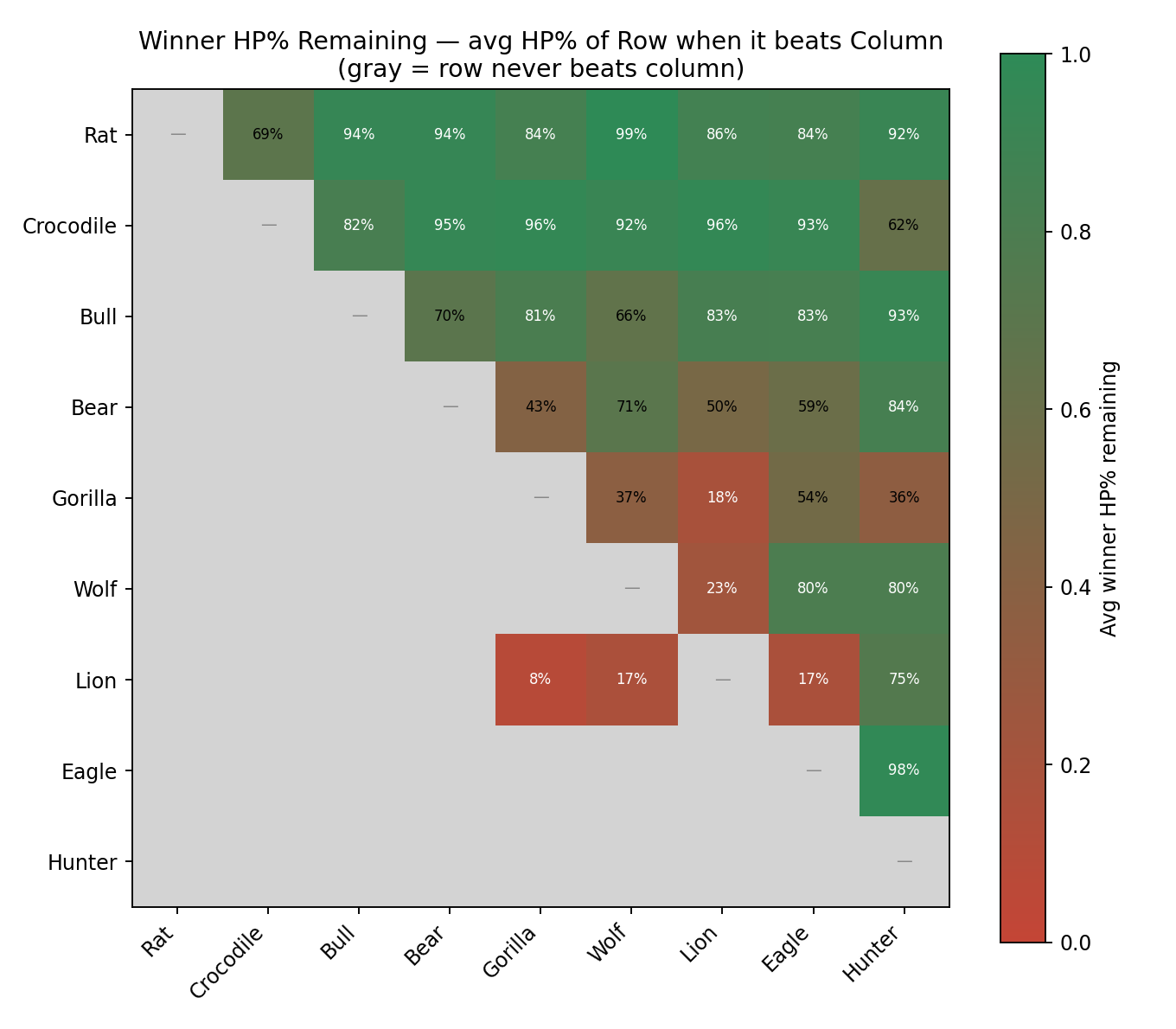}
    \caption{Pairwise HP-remaining matrix for group one-vs-one encounters. Each cell represents the percentage of HP remaining for each winner, from 500 Monte Carlo trials per matchup.}
    \label{fig:hp-matrix}
\end{figure}

The most significant shift is the emergence of swarm dominance. Rats, which perform poorly in isolation, overwhelmingly defeat all other species in group settings. Empirical results show that even high-damage units such as bears, crocodiles, and bulls are unable to eliminate rats quickly enough to offset the rate at which attackers accumulate, often leaving thousands of surviving rats at the end of an encounter (shown in Figure \ref{fig:hp-matrix}). More generally, outcomes are governed by scaling behavior: species with more units can apply damage more efficiently than species with higher individual damage but fewer units.

Among non-swarm species, a consistent hierarchical ordering remains. Crocodiles reliably defeat bulls, bears, and other large melee units due to superior durability and numbers, while bulls and bears maintain advantages over mid-tier species such as lions and wolves. In contrast, lower-tier and specialized units, including eagles and the hunter, fail to scale effectively; they are consistently overwhelmed in group combat. Overall, these results demonstrate that group interactions shift the governing dynamics from individual capability to scaling efficiency, largely mirroring our biomass predictions in Table \ref{tab:biomass_ranking}.

\subsection{Pick Two Results}

We now evaluate the full Pick Two scenario, in which two selected defender species must protect a central VIP against the remaining seven attackers. These trials capture the combined effects of coalition composition, multi-agent interaction dynamics, and emergent behavior under adversarial pressure.

\begin{figure}
    \centering
    \includegraphics[width=0.8\linewidth]{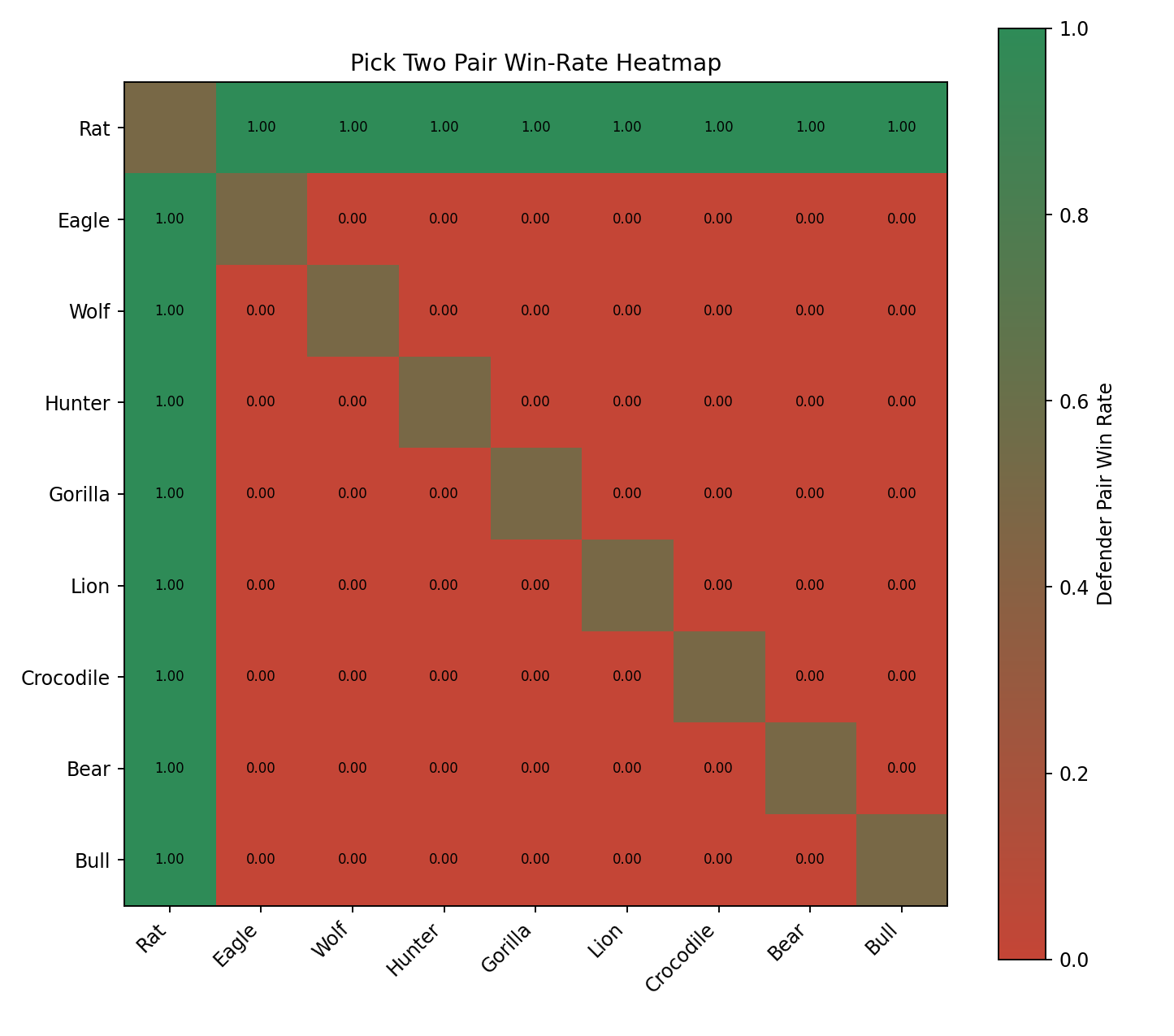}
    \caption{Pairwise win-rate matrix for Pick Two. Each cell represents the probability that the row species and the column species successfully protect the VIP, estimated over 500 Monte Carlo trials per matchup.}
    \label{fig:Picktwo-matrix}
\end{figure}

The Pick Two results (Figure~\ref{fig:Picktwo-matrix}) reveal a highly skewed performance landscape in which only a small subset of defender pairings achieve VIP survival. Most non-viable coalitions collapse within seconds (Figure~\ref{fig:PickTwoTimeMatrix}), often leaving the attacking force largely intact. These outcomes reinforce the group one-vs-one results, demonstrating that multi-agent combat strongly favors strategies that scale efficiently under numerical pressure.

A clear pattern is the central role of rats in all successful strategies. Every winning configuration includes rats as one of the two defenders, while pairings without rats consistently fail. When combined with rats, even weaker or non-scaling units become viable, demonstrating that survival is primarily governed by the ability to sustain large numbers of concurrent engagements rather than individual combat strength. However, the second defender still matters: high-durability units such as crocodiles or units with high effectiveness against swarms, like eagles or the hunter, increase the odds of protecting the VIP. Overall, the results highlight that effective coalitions emerge from favorable scaling rather than additive strength.

\subsection{Sensitivity Analysis}
To evaluate the robustness of the observed coalition dynamics, we performed a sensitivity analysis by uniformly reducing unit counts across all species. We tested count multipliers of 0.5x and 0.2x relative to the baseline configuration while holding all other parameters fixed. These settings reduce overall engagement capacity and allow us to examine whether swarm dominance persists with fewer units. Results can be seen in Figures \ref{fig:.5-Picktwo-matrix} and \ref{fig:.2-Picktwo-matrix}.

Reducing unit counts weakens swarm-based strategies, particularly those involving rats, but does not produce alternative successful coalitions. At 0.5x, rat-based pairings remain the only viable defenders, though with reduced reliability. At 0.2x, their effectiveness largely collapses, and win rates across all defender pairs approach zero. Importantly, non-swarm pairings do not improve under these conditions, indicating that reduced numerical pressure does not rebalance the system but instead degrades overall performance. These results suggest that sufficient engagement concurrency is a necessary condition for success, and that while swarm dominance depends on scale, the relative structure of coalition performance remains unchanged.

\section{Discussion}

\textbf{Analysis of Results:} 

The results demonstrate that unit numbers are the primary driver of combat outcomes in the simulated environment. While individual strength determines performance in isolated encounters, its influence diminishes substantially in multi-agent settings, where the ability to sustain many simultaneous engagements becomes the dominant factor. The dominance of smaller units in the simulation is likely not purely an artifact of the modeling framework but instead reflects several structural advantages associated with high-count groups in multi-agent systems. 

Large numbers allow many agents to apply damage simultaneously, while larger animals remain constrained by attack cadence, turning rate, and target switching, creating sustained engagement pressure that favors high-count groups. High-count groups also saturate the arena by surrounding opponents and limiting maneuverability, increasing sustained contact time, and making disengagement difficult. More broadly, these outcomes are consistent with action-economy effects observed in other multi-agent systems, where the number of independent attackers can outweigh individual unit strength. That said, the simulation may amplify these dynamics through modeling assumptions that favor sustained engagement concurrency, including flat terrain, simplified collision dynamics, the absence of morale or exhaustion, and limited area-of-effect or trampling behaviors. 

Consequently, the results should not be interpreted as literal predictions of real-world animal combat outcomes, but rather as evidence that under sustained multi-agent pressure, numerical scaling and engagement concurrency can dominate individual combat strength.

\textbf{Answering Research Questions:}
\begin{enumerate}
    \item \textit{How is survival performance distributed across all possible defender pairs?}

    Survival performance is highly uneven across the space of defender pairs. Only a small subset of coalitions consistently protect the VIP, with successful strategies overwhelmingly involving units capable of sustaining large numbers of simultaneous engagements. In particular, rat-based coalitions dominate the performance landscape, while most non-scaling pairings fail rapidly under numerical pressure.

    \item \textit{Do certain coalitions exhibit disproportionately high or low performance relative to the capabilities of their individual components?}

    Coalition performance cannot be predicted from individual capability alone. Instead, outcomes depend strongly on how species scale in multi-agent settings and interact within a coalition. Pairings that combine high-count species with durable or high-damage defenders consistently outperform other combinations, while coalitions composed of individually strong but poorly scaling species often fail. These results suggest that coalition effectiveness emerges primarily from complementary scaling behavior rather than individual strength alone.

    \item \textit{How sensitive is performance to parameter adjustment?}

    Sensitivity analysis indicates that while absolute performance degrades as overall unit counts are reduced, the underlying structure of coalition effectiveness remains largely unchanged. Rat-based strategies weaken under reduced counts, but alternative non-scaling coalitions do not emerge as viable replacements. This suggests that the dominance of scaling behavior is a robust feature of the system rather than an artifact of a particular parameter setting, though its precise manifestation may depend on environmental and modeling assumptions not varied in the present study.
\end{enumerate}

\textbf{Limitations:} 
This study relies on several simplifying assumptions that may influence the observed outcomes. First, agent behavior is modeled using a small set of fixed archetypes, which capture broad strategic patterns but do not account for learning, adaptation, or more nuanced decision-making that may arise in real-world interactions. Second, damage values are derived from calibrated proxies such as bite force and momentum, introducing uncertainty due to limited cross-species comparisons and the need for simplifying scaling relationships. While these choices provide a consistent and interpretable framework, they may not fully reflect the true variability in combat effectiveness. Finally, the simulation is conducted in a simplified, flat environment without terrain or obstacles. This removes potential influences such as chokepoints, visibility constraints, and mobility restrictions, which could significantly alter engagement dynamics and the relative effectiveness of different strategies. Together, these limitations suggest that while the results capture key structural behaviors of the system, their quantitative outcomes should be interpreted within the context of these modeling constraints.

\textbf{Reproducibility}
The simulation framework is available and runnable as part of the project artifact. The release includes the executable build, experiment scripts, analysis scripts, sample outputs, and a README file explaining how to reproduce the one-vs-one, Pick Two, sensitivity, and figure-generation results.

Experiments are executed through provided PowerShell scripts that automate shard-based simulation runs, data collection, and merging. The Python analysis scripts can then be run to regenerate the reported figures from the merged CSV outputs. Fixed experiment seeds are used to support reproducible trial outcomes and aggregate performance metrics.

The artifact package and instructions are available at \url{https://jcvanly.github.io/}.

\textbf{Future Work:} 
Several extensions could improve both the realism and generality of the current model. Incorporating adaptive behaviors, such as learning-based or state-dependent policies, would allow agents to modify their strategies in response to changing conditions rather than relying on fixed archetypes. Introducing more realistic environmental structure, including terrain variation, obstacles, and visibility constraints, could reveal how spatial factors influence engagement dynamics and coalition effectiveness. Expanding the set of modeled species would further test the robustness of the observed scaling effects and provide a broader basis for comparison. Finally, exploring alternative scenarios—such as different objective functions, asymmetric starting conditions, or varying attacker-defender ratios—would help determine how sensitive the current findings are to the specific structure of the Pick Two framework.
\section{Conclusion}

In this work, we investigated the Pick Two problem through large-scale simulation, analyzing how individual capability, group dynamics, and coalition structure influence survival outcomes. Our results demonstrate a clear shift from individual strength to scaling efficiency as the dominant factor in multi-agent settings. While one-on-one interactions produce an interpretable hierarchy based on per-unit attributes, group interactions amplify small advantages, resulting in highly polarized and largely deterministic outcomes.

Across all experiments, effective performance is concentrated in a small subset of strategies. In particular, swarm-based agents dominate due to their ability to sustain large numbers of concurrent engagements, overwhelming opponents whose damage output remains sequential. In the full Pick Two scenario, coalition effectiveness is not additive: successful defender pairs emerge from complementary scaling behaviors, while combinations of individually strong agents frequently fail. These findings highlight the importance of interaction structure and parallelism in determining system-level outcomes.

Sensitivity analysis shows that these dynamics are robust to moderate changes in model parameters. Although variations in unit count can shift the relative strength of specific strategies and introduce threshold effects, the overall pattern persists: strategies that maximize engagement capacity consistently outperform those that rely on individual strength alone.

Overall, this study demonstrates that the Pick Two problem serves as a useful testbed for understanding broader principles in multi-agent systems. In particular, it illustrates how collective behavior, concurrency, and scaling effects can dominate over individual capability. Future work may extend this framework by incorporating more complex environments, richer agent behaviors, and alternative interaction models to further explore the role of structure in coalition performance.

% ===== Appendix =====
\appendix
\section{Additional Data on Flight Software Dependencies}
\section{Appendix}
\label{sec: Appendix}

\begin{figure}[H]
    \centering
    \includegraphics[width=0.8\linewidth]{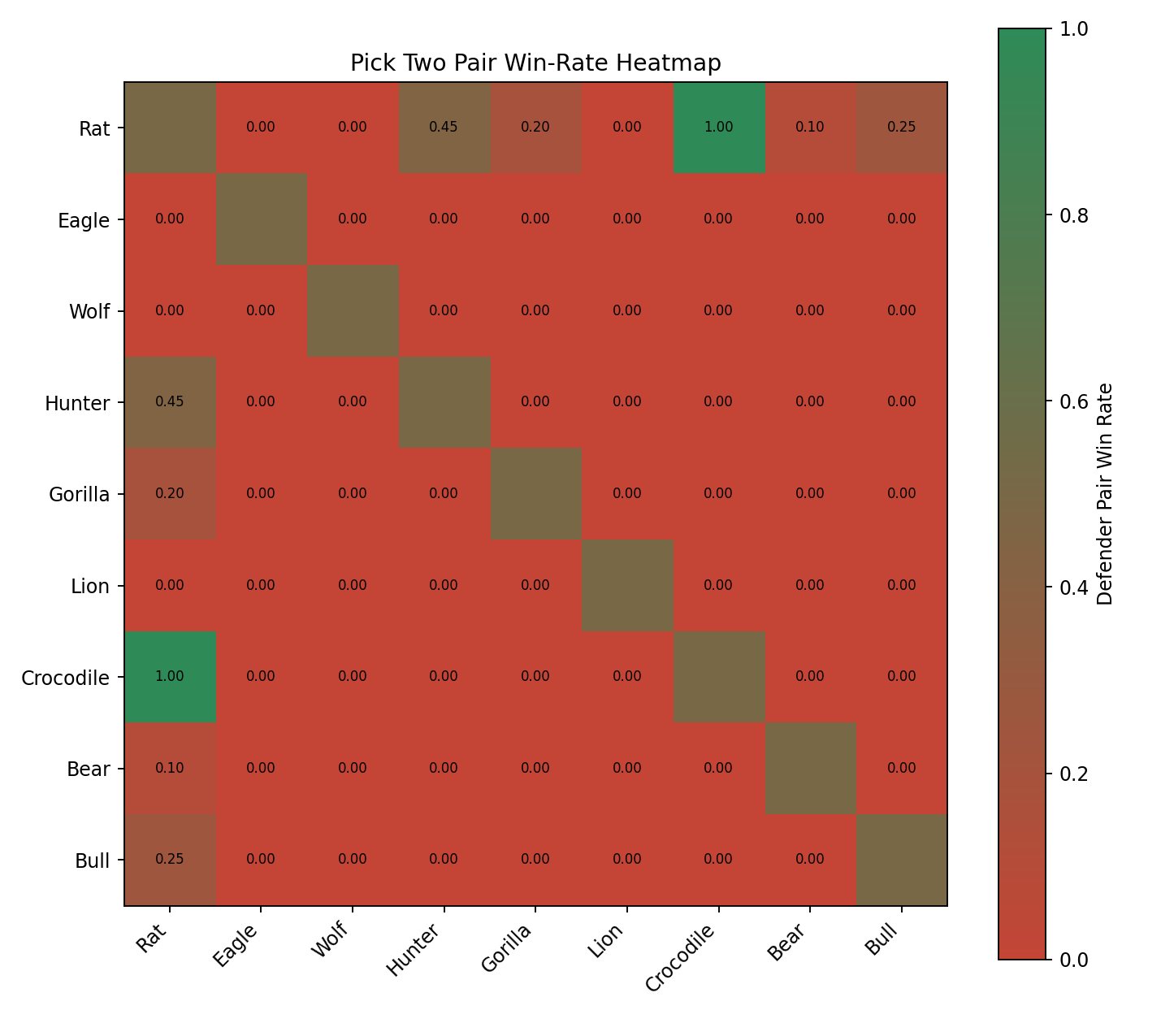}
    \caption{Pairwise win-rate matrix for Pick Two at 0.5x scale. Rat based coalitions are significantly less effective than with full numbers. Estimated over 500 Monte Carlo trials per matchup.}
    \label{fig:.5-Picktwo-matrix}
\end{figure}

\begin{figure}[H]
    \centering
    \includegraphics[width=0.8\linewidth]{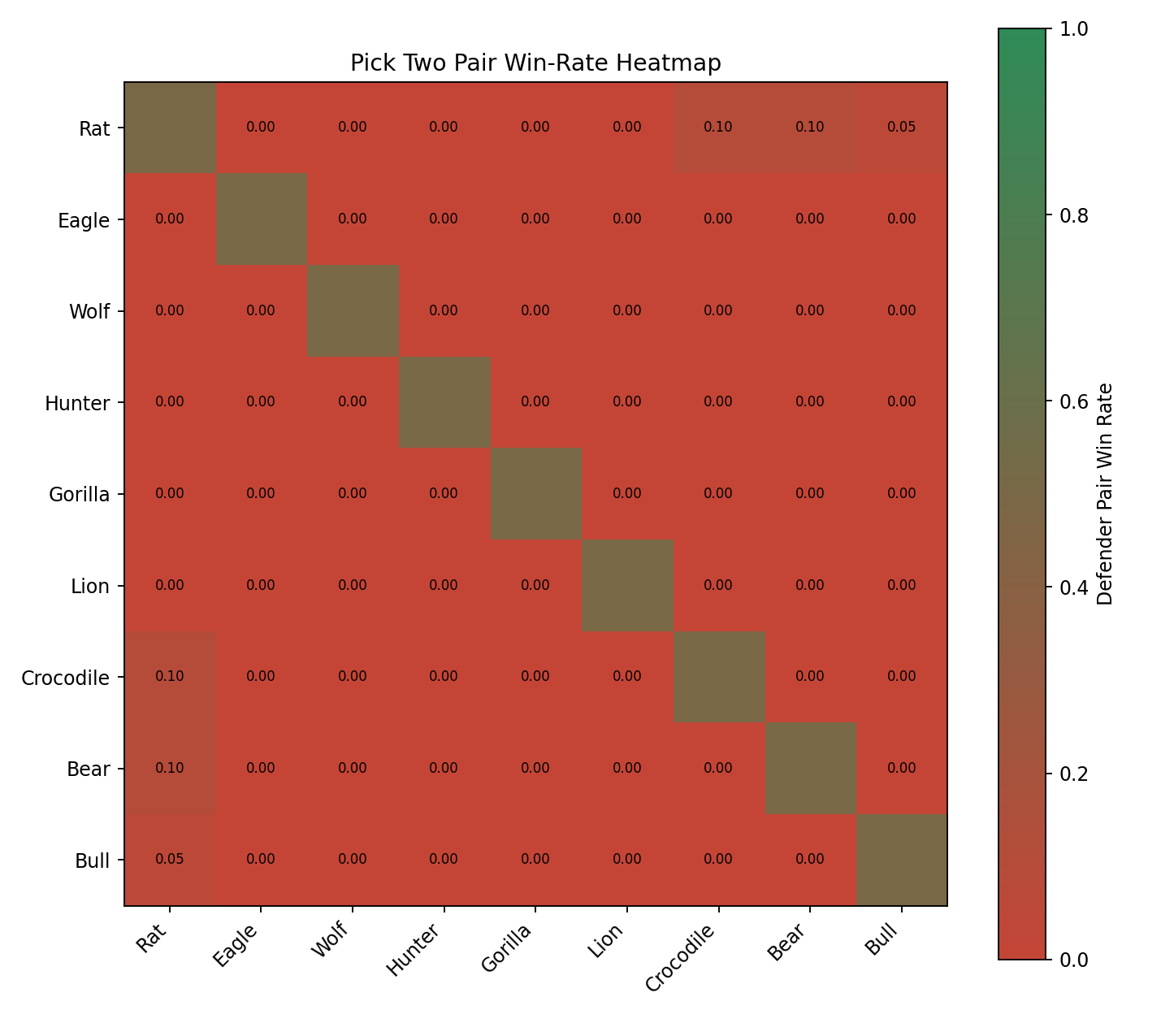}
    \caption{Pairwise win-rate matrix for Pick Two at 0.2x scale. Almost all coalitions fail to protect the VIP demonstrating the power of scaling. Estimated over 500 Monte Carlo trials per matchup.}
    \label{fig:.2-Picktwo-matrix}
\end{figure}

\begin{figure}[H]
    \centering
    \includegraphics[width=0.8\linewidth]{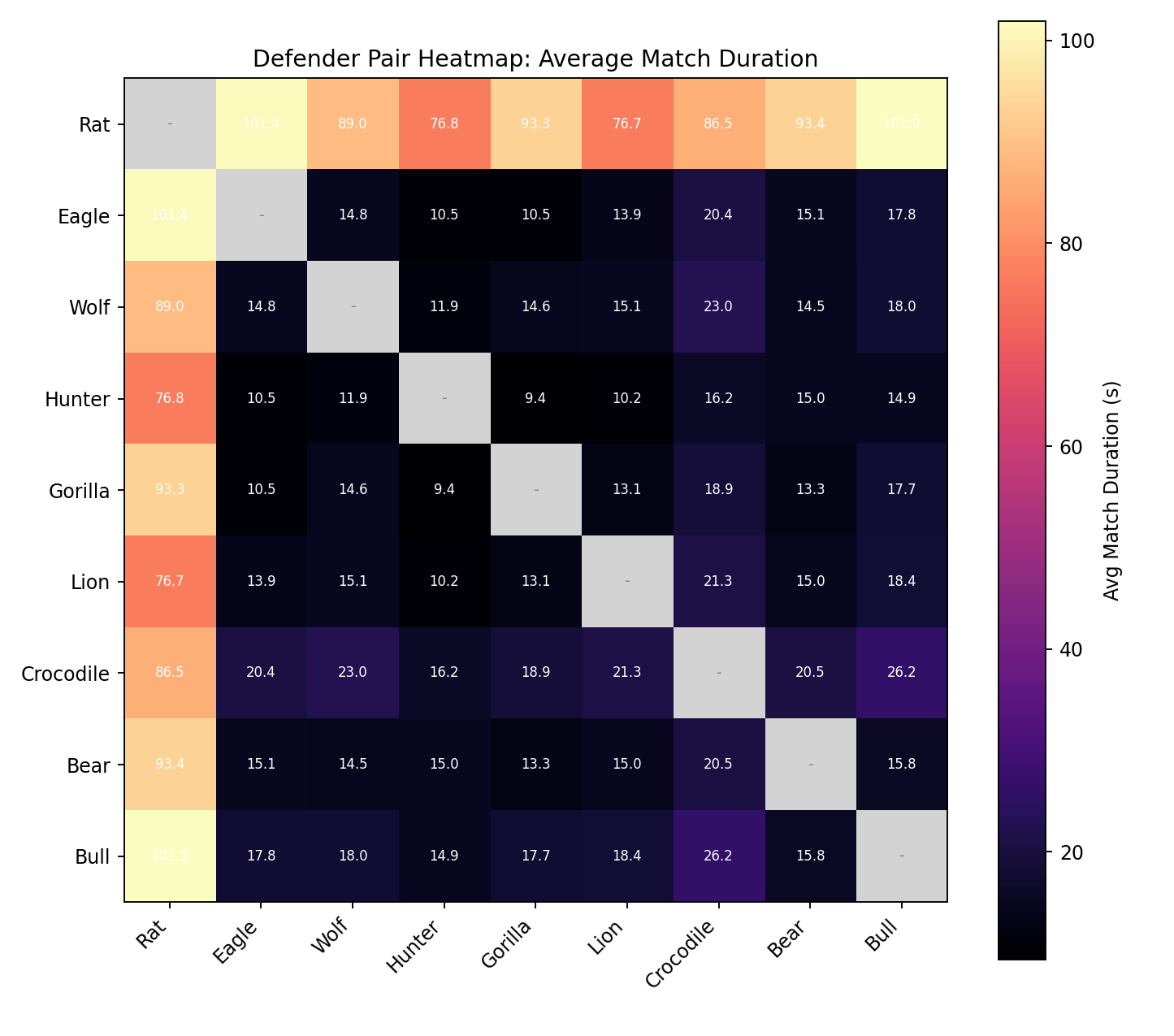}
    \caption{Pairwise match time matrix for Pick Two. Each cell represents the time that the row species and the column species successfully protect the VIP or are all eliminated, estimated over 500 Monte Carlo trials per matchup.}
    \label{fig:PickTwoTimeMatrix}
\end{figure}

\bibliographystyle{IEEEtran}
\bibliography{sources}

\end{document}